\documentclass{sig-alternate}
\usepackage{colortbl}
\usepackage{ctable}
\usepackage{tabularx}
\usepackage{slashbox}
\usepackage{subfigure}
\usepackage{algorithm}
\usepackage{algpseudocode}
\usepackage{pseudocode}

\newcommand{\MyCaption}[3]{\caption[#1]{#1 #2\label{#3}}}

\newcommand{\MyEquation}[1]{\begin{equation}
\ensuremath{\input{Equations/#1}}
\end{equation}
}

\newcommand{\MyEquationInline}[1]{\ensuremath{\input{Equations/#1}}}

\newcommand{\MyEquationLabeled}[2]{\begin{equation}
\ensuremath{\input{Equations/#1}}
\label{#2}
\end{equation}
}

\newcommand{\MyEquationReference}[1]{(see Equation \ref{#1})}

\newcommand{\MyFigureReference}[1]{(see Figure \ref{#1})}

\newcommand{\MyIgnore}[1]{}

\begin{document}
%
\conferenceinfo{HT'10,} {June 13--16, 2010, Toronto, Ontario, Canada..}
\CopyrightYear{2010}
\crdata{978-1-4503-0041-4/10/06}
\clubpenalty=10000
\widowpenalty = 10000

\title{Analysis of Graphs for Digital Preservation Suitability}
\numberofauthors{2}
\author{
\alignauthor
Charles L. Cartledge\\
       \affaddr{Old Dominion University}\\
       \affaddr{Department of Computer Science}\\
       \affaddr{Norfolk,VA 23529 USA}\\
       \email{ccartled@cs.odu.edu}\\
\alignauthor Michael L. Nelson\\
       \affaddr{Old Dominion University}\\
       \affaddr{Department of Computer Science}\\
       \affaddr{Norfolk, VA 23529 USA}\\
       \email{mln@cs.odu.edu}\\
}
\date{\today}

\maketitle
\begin{abstract}
We investigate the use of autonomically created small-world graphs as a
framework for the long term storage of digital objects on the Web in a
potentially hostile environment.  We
attack the classic Erdos --- Renyi random, Barab{\'a}si and  Albert
power law, Watts --- Strogatz small world and our Unsupervised
Small World (USW) graphs using different attacker strategies and report their
respective robustness.  Using different attacker profiles, we construct a game
where the attacker is allowed to use a strategy of his choice to remove a
percentage of each graph's elements.
 The graph is then allowed to repair some portion of its self.  We report on the
number of alternating attack and repair turns until either the graph is
disconnected, or the game exceeds the number of permitted turns.  Based on our
analysis, an attack strategy that focuses on removing the vertices with the highest
\emph{betweenness} value is most advantageous to the attacker. Power law graphs can
become disconnected with the removal of a single edge;
random graphs with the removal of as few as 1\% of their vertices, small-world
graphs with the removal of 14\% vertices, and USW with the removal of
17\% vertices.  Watts --- Strogatz small-world graphs are more robust and
resilient than random or power law graphs.  USW graphs are more
robust and resilient than small world graphs.  A graph of USW connected WOs filled with data
could outlive the individuals and institutions that created
the data in an environment where WOs are lost due to random failures or directed attacks.
\end{abstract}
\category{H.4}{Information Systems Applications}{Miscellaneous}
\category{I.6.8}{Simulation and Modeling}{Types of Simulation}
\category{E.m}{Data}{Miscellaneous}
\terms{Algorithms, Experimentation, Reliability, Theory}
\keywords{small world, robustness, resilience}
\section{Introduction}
We are exploring the creation of web objects (WO) that establish
and maintain links between themselves and can live without the
intervention of conventional repositories.  We are investigating how data
inside the WOs could outlive the individuals and institutions that created
the data with little or no external guidance or direction making them ideal for
use in environments where the stewardship of the data is paramount.  WOs can be thought of as
having all the properties Kahn-Wilensky Framework digital objects
 \cite{kahn2006fdd} except that they live directly in the Web Architecture
 \cite{Jacobs:2004ul} and do not require an explicit repository system
(e.g., DSpace, Fedora) for management, nor do they require global
knowledge of the entire network.

These WOs send messages back and forth between each other and
have the desirable small-world characteristics of relatively high clustering
coefficients (where there is a high probability that ``a friend of a friend is
my friend as well'') and a short average path length between any two nodes in
the graph.  We have developed an Unsupervised Small World (USW) algorithm that
creates graphs with the desired characteristics by adding one node
at a time to an existing graph.  Edges between nodes are created based on local
information that each new node discovers from nodes already part of the graph. A
graph of these WOs is robust and individual WOs can communicate with one another
even when the underlying infrastructure has been damaged or disabled using their
internal and locally maintained data structures.  Some of the messages that the
WOs could exchange might include the location of a service that wold migrate
data in an ``old'' format to a ``new'' one (migration from GIFF to JPEG), the
location of a new server willing and ready accept additional WOs for storage
(refreshing of the bits), etc. to support preservation efforts.

This is an extension of  our prior work on self-contained digital
objects  \cite{374342} and earlier investigations into the creation
of self-arranging networks  \cite{1378990}.  In this paper we present
the simulation results of an algorithm for digital objects to create
a small-world Graph without direct supervision by an administrator or
repository.  Although assistance from administrators or repositories
is possible in this model, it is not required.  The motivation for and
scenarios of how a network of WOs could engage in digital preservation
tasks is covered in  \cite{jcdl09-usw}; this paper presents only the
analysis of an algorithm to test the robustness and resilience of such a network.

Milgram  \cite{milgram1967small} is credited with formulating the idea that in a
social network, the path length between any two randomly selected individuals in
the US in the 1960 was on average between 5 and 6, leading to the phrase ``6
degrees of freedom.''  An algorithmic technique from a totally ordered $k$-degree lattice
to a random graph passing through a phase that exhibited
Milgram's small-world characteristics was made popular by Watts --- Strogatz
  \cite{watts:collective_dynamics}.  Their technique required a $k$ degree lattice
as a foundation before a small-world could be constructed.

We would prefer if the WOs could self-organize into a graph
exhibiting small-world properties without first creating a regular
or random graph as a starting point.  Small-world graphs are
interesting because they occur frequently in a variety
of different fields.  They have been found in cellular metabolism,
Hollywood actor relationships, Internet routers, protein regulatory
networks, research collaborations, sexual relations and World Wide
Web page linkages  \cite{barabasi:scale_free_networks}.
 Furthermore, current methods for
small-world graph creation are based on an outside view of the network
and an omnipresent/omnipotent view of the graph structure.

We test the robustness and resilience of classical random, power law, small-world
 and our USW graphs by subjecting each to a set of
strategies that an attacker might use to disconnect the graph.  We test these
graphs by using a game where the attacker and the graph alternate turns.  The
attacker will be able to inflict damage on the graph (as a test of the graph's
robustness), after which the graph will be able to repair and strengthen itself
(as a test of the graph's resilience).  Our efforts are focusing on the
performance of the graphs during different stress tests.  We are
primarily interested in the autonomic processes that create and preserve the
graph.  By expanding the contents of the node from purely maintenance data to
payload data (files, images, or other binary information), the contents of a USW
graph could be preserved from loss even in the face of repeated censorship
attacks.
\section{Types of Graphs Based on Degree Distributions} \label{sec:typeOfGraphs}
Graphs can be classified by many different and overlapping criteria including
the presence or absence of well defined structural elements.  Randi{\'c} and
DeAlba  \cite{randic1997dense} provide an extensive list of different
classifications.  Within this paper, we are interested in the classifying graphs
by their degree distributions.  Those processes can be purely random,
power law, classical Watts --- Strogatz small-world, or our
USW construction process.

Each of these processes generates a graph with distinctively different degree
distributions, clustering coefficients (CC) and expected average path
lengths.  Figure \ref{fig:comparison} is a plot of representative degree
distributions for each of these types of graphs.  In
Figure \ref{fig:comparison}, the red
circles are characteristic of a power law distribution.  The
black x's are from a small-world graph and look very much
like a random distribution because the underlying methodology
for creating the edges is random.  The difference between a
small-world distribution and a random one is the smallness of
the degree distribution $\sigma$ and having a mean $\mu$ that is same as the
underlying
lattice that was used as the base.  While this small-world distribution is $\pm$
4, a similar random one is $\pm$ 10.  A random graph degree distribution is
shown with the green triangles, whose $\mu$ is centered at $p*n$ and a
Poisson distribution for the rest of the degreed nodes.  The USW
construction parameters will affect the center of the
blue crosses and where that center lies on the x-axis.  USW
parameters $\beta$ and $\gamma$ were set to 0.95 each in order
to separate the USW graph from the other graphs.  USW $\gamma$
affects the left-right location of the center, while $\beta$
affects the height of the center. The CC for random
and small-world graphs are \MyEquationInline{clusteringCoefficientRandom} and
\MyEquationInline{clusteringCoefficientSmallWorld} respectively.  While the
average path lengths are \MyEquationInline{averagePathLengthRandom} and
\MyEquationInline{averagePathLengthSmallWorld} respectively
 \cite{watts:collective_dynamics}.  The CC and average path length of power law
 and USW graphs are not tractable and do not have a
closed form solution.

\begin{figure}
 \centering
\includegraphics[width=2.35in,angle=-90]
{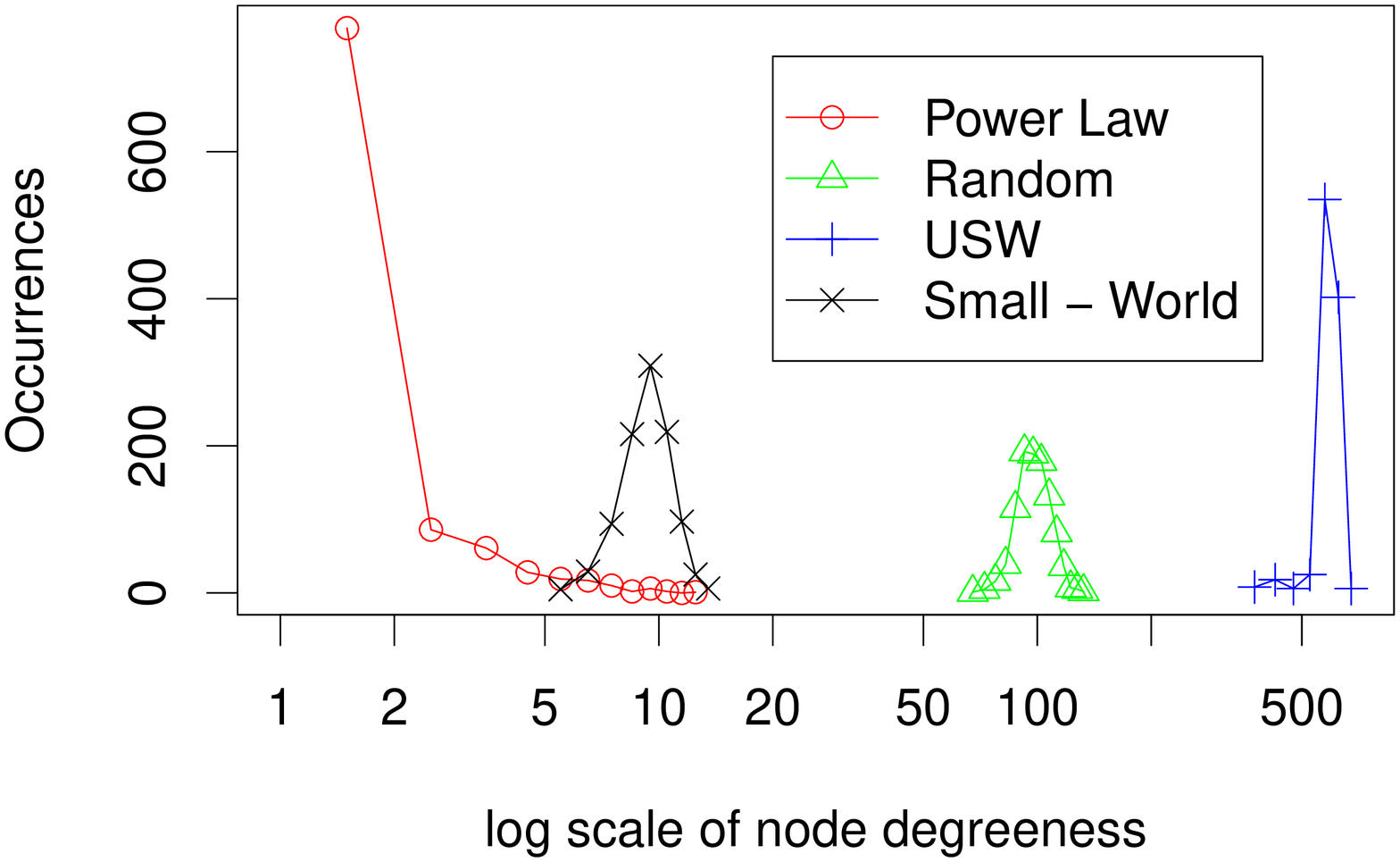}
\MyCaption{Histogram of representative degree distributions of 1000 node graphs built using
random, power law, small-world and USW
processes.}{}{fig:comparison}
\end{figure}

\subsection{Random}
A random  graph is one that is generated by some random
process  \cite{janson2000random, baumann2005network}.  These graphs can be created
by many non-equivalent techniques.
At the end of the random graph construction process, the graph may not be connected.
\subsection{Power law}
Power law graphs are characterized by the fraction of their vertices that have a
specified degree $k$. In general, the degree distribution of a power law graph
is given by:
\MyEquationInline{pa}.
Preferential  attachment graphs are a special case of \emph{Power Law} graphs.  Preferential attachment graphs grow over time by the addition of new vertices.
\subsection{Small-world}
Small-world graphs introduced in
 \cite{watts:collective_dynamics,newman-2000-101} begin with a $k$-lattice
and rewiring each edge with a probability $p$.  Small-world graphs may be
planar or non-planar and there is a greater than 0 probability that the
resulting graph will be neither simple and nor connected.
Small-world graphs have distinctive average path length and
clustering coefficient properties.

The  average distance between any two vertices is small (growing
logarithmically with the number of vertices in G). The average clustering
coefficient for the graph is high.  A clustering
coefficient is the fraction of vertices that any two vertices have in
common over the complete set of adjacent vertices that the two original vertices
have. Random graphs have low average path distances, but
their average clustering coefficients tend towards 0.
\subsection{Unsupervised Small World}
Unsupervised Small World (USW)  graphs are simple, connected non-planar built by
the autonomous actions of each node as it is added to the existent graph.  USW
graphs are characterized by a clustering coefficient and average path length
comparable to that of traditional small-world graphs, but markedly different in
how they are constructed. The autonomous
algorithm that each independent node uses to locate itself within, and thereby
grow the USW is detailed in  \cite{jcdl09-usw,1378990}.

\section{Attacker profiles}
The selection of an attack profile involves is a decision based on many
different pieces of data.  Included in this list of data is:
\begin{itemize}
 \item  The goal of the attack (for instance complete removal of the graph, or
inflicting enough damage to the graph to cause the graph to become
disconnected),
\item Which metric to use to measure the progress of the attack,
\item What type of graph component (edge or vertex) to remove,
\item  What technique to use to select the most ``central'' or ``vital''
component of the graph to remove,
\item  When the components are ranked based on their centrality, which specific
one to select.
\end{itemize}
\subsection{Attack progress metrics}
The following metrics will be collected because they relate directly to the
ability of the graph to communicate effectively between its WOs, and will change
depending on the amount of damage the graph sustains.

\textbf{Average inverse path length} is the inverse of the mean of all the
shortest paths in the graph.  Because the shortest path between vertices in two
different components is $\infty$, the inverse is 0 and therefore is a valid
value
that does not cause the computation to fail.  A larger average inverse path
length means that the distance between nodes is on average shorter
 \cite{holme2002attack}. \MyEquation{averageDistanceInverse.tex}

\textbf{Average path length} is the mean of the all the shortest (geodesic)
paths
in the graph \MyEquation{averageDistanceConnected}

\textbf{Clustering Coefficient} is the
likelihood that two neighbors of $v$ are connected
 \cite{newman2003structure,opsahl2009clustering}
\MyEquation{clusteringGlobal}

\textbf{Density} is the ratio for the edges and nodes that are members of the
connected graph  \cite{randic1997dense}\MyEquation{density2}

\textbf{Damage} is the ratio of the largest component to the entire graph
 \cite{odu:techreport:2010}
\MyEquation{damage}

\textbf{Diameter} is the maximal shortest path between any vertices $u$ and $v$
\MyEquation{diameter}

Where $n$ is the number of nodes in the entire graph, $m$ is the number of edges
in the entire graph, $p$ is a probability of connection or rewiring (based on
the type of graph).

During the course of the simulation, we expect the following changes in each of
the above metrics when the graph conducts maintenance:
\begin{description}
\item [Average inverse path length] to remain nearly the same or to growly
slowly as the graph becomes more and better connected,
\item [Average path length] to decrease as more alternative paths are created,
\item [Clustering Coefficients] to increase as more triads are created because
of the increasing edges,
\item [Density] to increase because the graph creates more edges,
\item [Damage] to remain nearly static, and
\item [Diameter] to decrease as the graph becomes more and better connected.
\end{description}

\subsection{Centrality measurements}
A  centrality measurement is a way of quantifying the notion that some
components of a graph are more important than others.  Some centrality
measurements are based purely on data that is available at the graph component
level and are invariant with respect to the rest of the graph; these are called
\emph{local} centrality measurements.  Other measurements are dependent on the
structure of the graph in to-to.  These are called \emph{global} centrality
measurements.
The difference between local and global knowledge is fundamentally one of degree
using the idea of $k-neighborhood$.  In the minimal case where $k=1$,
all knowledge is based on edges and vertices that are 1 edge away.  In the
maximal case where $k=D(G)$, all knowledge is based on total knowledge
of the graph.  Values of $k$ from 1 and $D(G)$ reflecting increasing knowledge
of G.
\subsubsection{Betweenness}
Betweenness is a global centrality measurement.  Betweenness is a measure of how
many geodesic paths from any vertices $s,t \in V$ use a either an edge
\MyEquationReference{eq:betweennessEdge} or a vertex \MyEquationReference{eq:betweennessVertex}. Removal of a graph component based on its
betweenness is a direct attack on the global structure of the graph.
\MyEquationLabeled{betweennessEdge}{eq:betweennessEdge}
\MyEquationLabeled{betweennessVertex}{eq:betweennessVertex}
\subsubsection{Closeness}
Closeness is a global centrality measurement.  Closeness quantifies the idea
that a vertex has a shortest average geodesic distance when compared to all
geodesic distances.
\MyEquationLabeled{closeness}{eq:closeness}
\subsubsection{Degreeness}
Degreeness is a local centrality measurement.  Degreeness is the number of edges
that are incident to a vertex \MyEquationReference{eq:degreeUndirectedDegree}.
Degreeness only makes sense for vertices. A vertex with a high degreeness is
central to a local portion of the graph, but not necessarily to the graph in to-to.
\MyEquationLabeled{degreeUndirectedDegree}{eq:degreeUndirectedDegree}
\subsection{Extremal values}
Selection  of an attack profile is dependent on an appropriately selected
centrality measurement, the type of graph component (either $E$ or $V$) to
delete and which of these components to delete based on the centrality
measurement selected.  Any centrality measurement will result in an unordered
list of numerical values.  Depending on the type of measurement selected, either
of the extremal values of \textbf{H}ighest or \textbf{L}owest will result in the
most disruption to the graph.
\subsection{Sample profiles}
Attacker  profiles are described with a three character token created from
permuting three disjointed sets.  The sets are: \{D,B,C\} representing the
centrality measurements \textbf{D}e\-greeness, \textbf{B}etwenneess and
\textbf{C}loseness; \{E,V\} representing graph component to be removed
\textbf{E}dge and \textbf{V}ertex, and \{L,H\} representing which extremal
value to use when selecting the component to be removed \textbf{L}ow and
\textbf{H}igh.  An example of one permutation is: \textbf{B-E-L} meaning that
betweenness centrality measurement is being computed, an edge will be selected
for removal and the lowest valued edge will be removed.

Each of the  different attack profiles is presented with the same graph
{\MyFigureReference{fig:sample}.  The attack profile continues to execute
until the graph is
disconnected.  In those cases where
there are multiple graph components with the same value (vertices of the same
degreeness, edges with the same betweenness, etc.), the attack profile is
recursively applied and the total number of deletions is reported.  Figure
\ref{fig:first} shows the sample graph is prior to the deletion of the first
attack profile specific element. Each attack profile assumes that the attacker
has complete (i.e., global) knowledge of the graph and so is able to make
decisions that are most advantageous to the attacker.  How this knowledge is
obtained is
outside this discussion.  The goal of each attack profile is the disconnection
of the graph, where disconnection is defined as the inability of vertex $i$ to
send a message to vertex $j$ \MyEquationInline{notallI2J}.
Therefore a graph with only one vertex is still connected and that removing a
vertex that is connected to only one other vertex does not disconnect the graph.

\newcommand{\MySize}[0]{1.95in}

\begin{figure*}
 \centering
\subfigure[B-E-L] {\includegraphics[width = \MySize, angle = -90]
{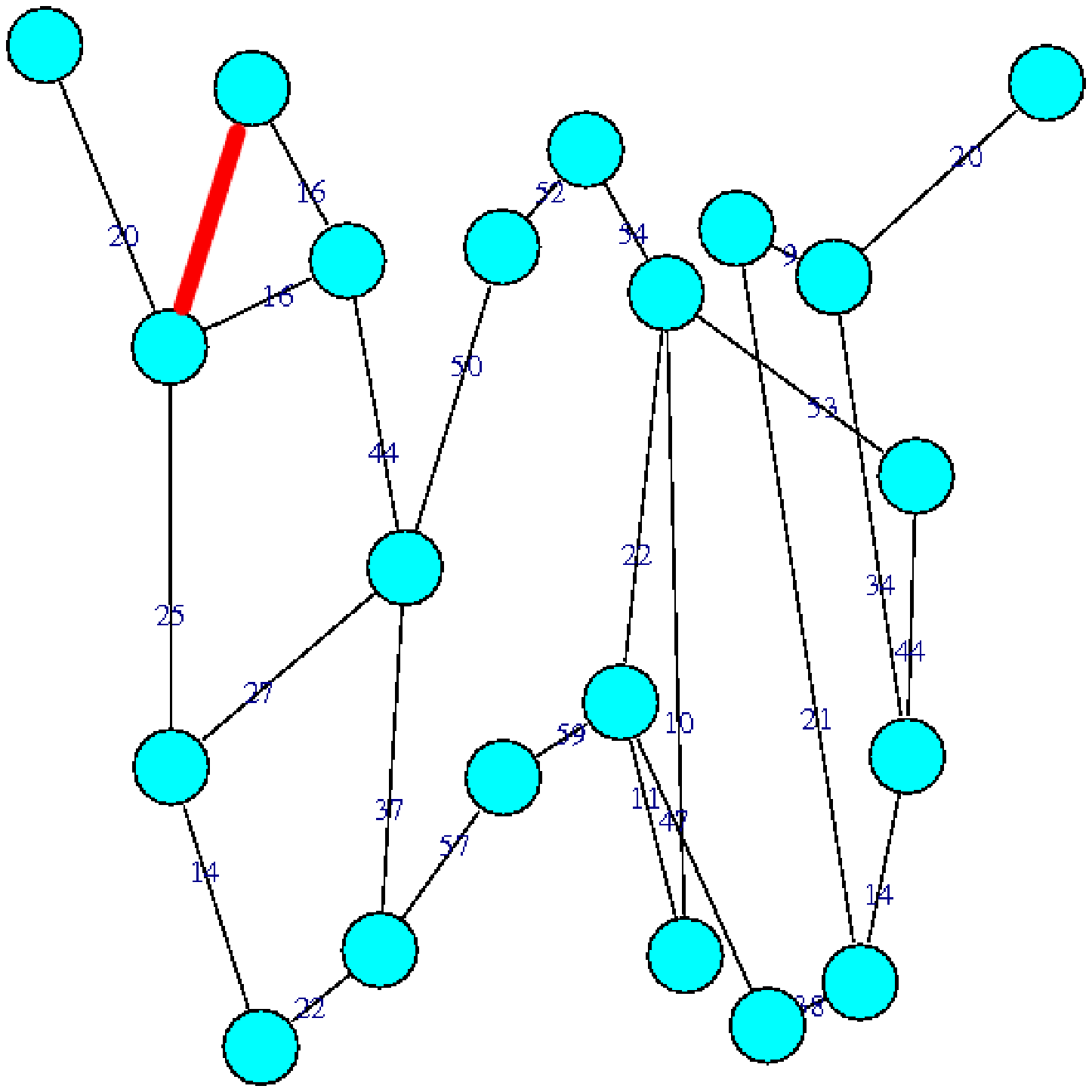}}
\subfigure[B-E-H] {\includegraphics[width = \MySize, angle = -90]
{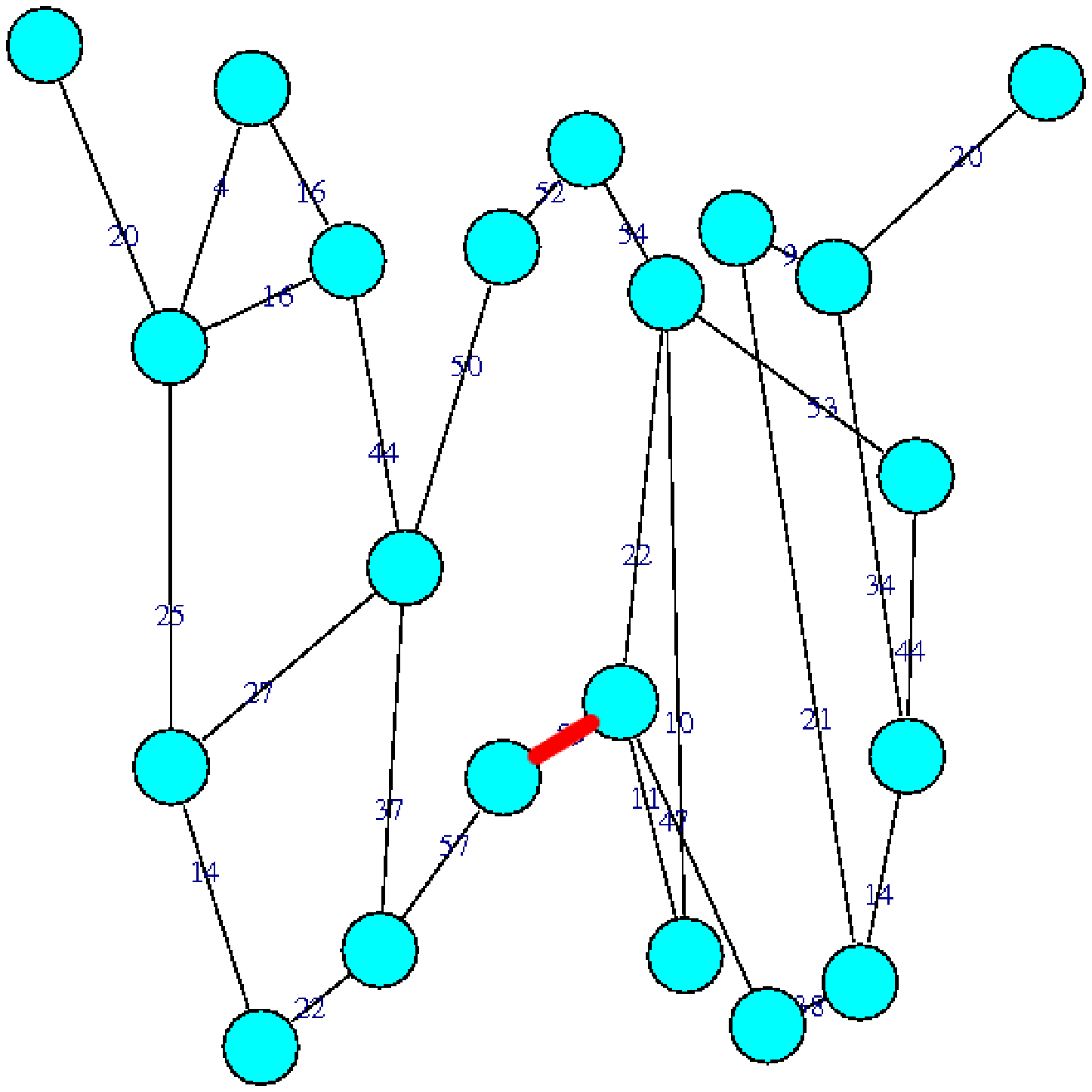}}
\subfigure[B-V-L] {\includegraphics[width = \MySize, angle = -90]
{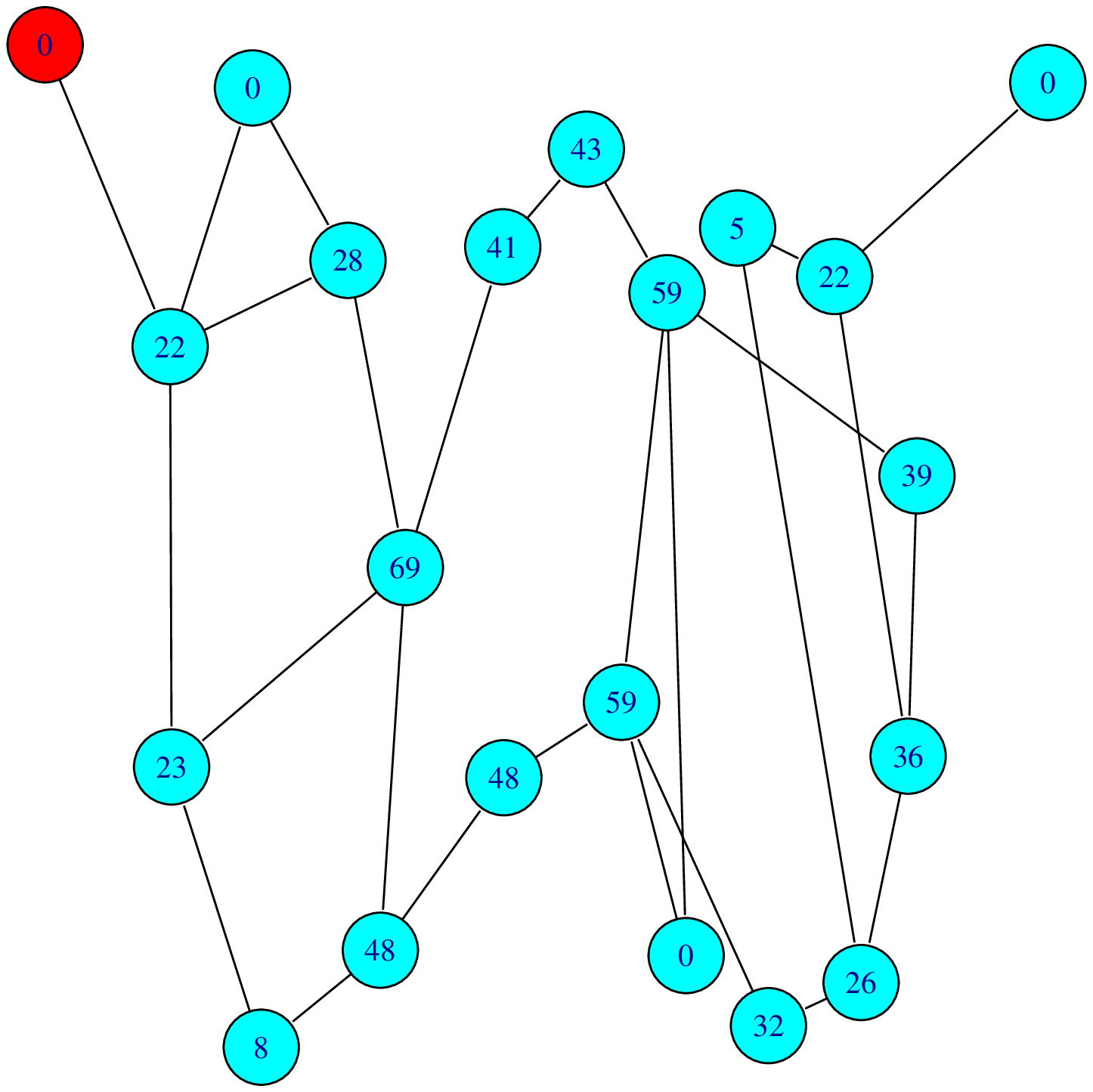}}
\subfigure[B-V-H] {\includegraphics[width = \MySize, angle = -90]
{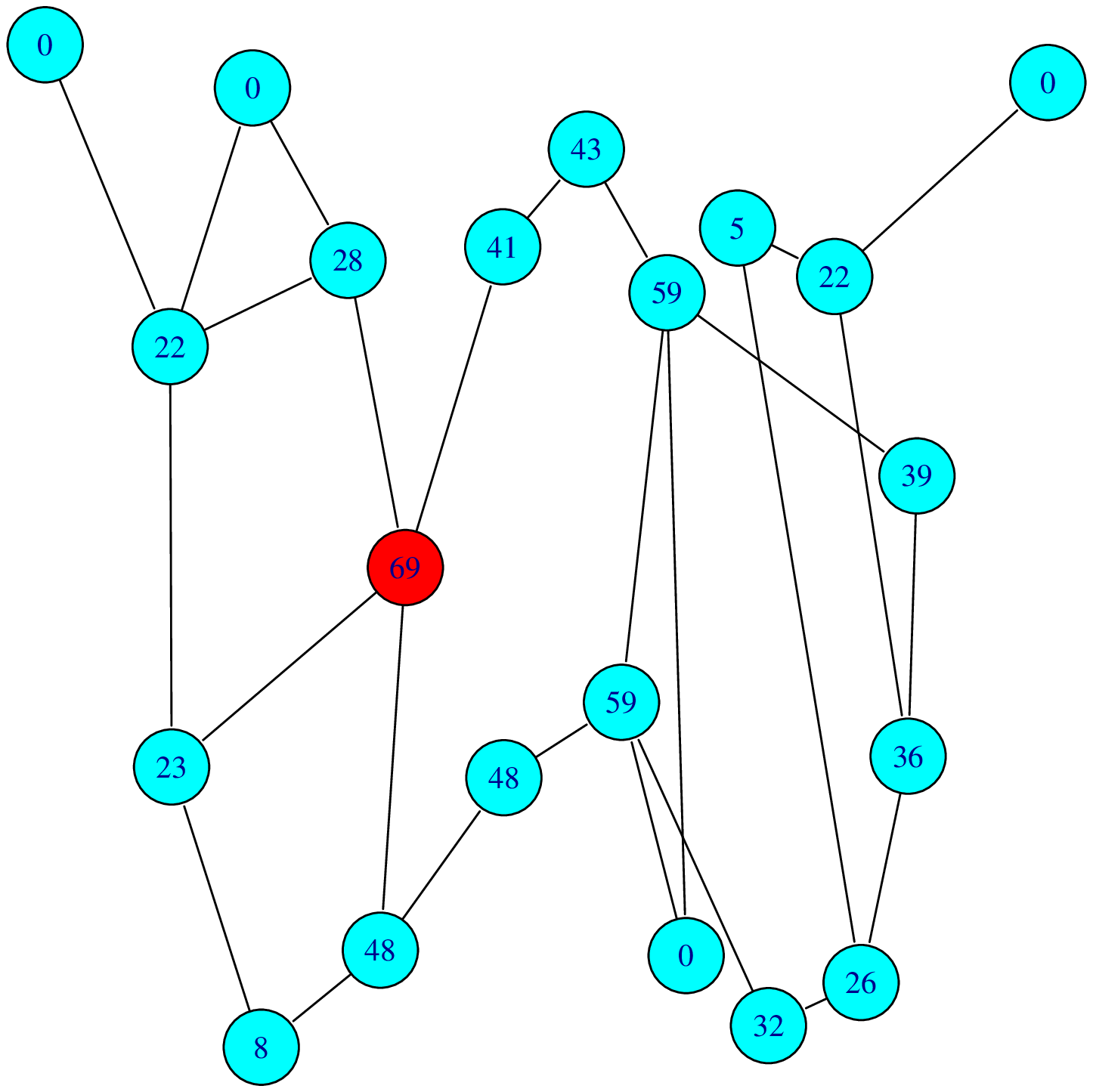}}
\subfigure[D-V-L] {\includegraphics[width = \MySize, angle = -90]
{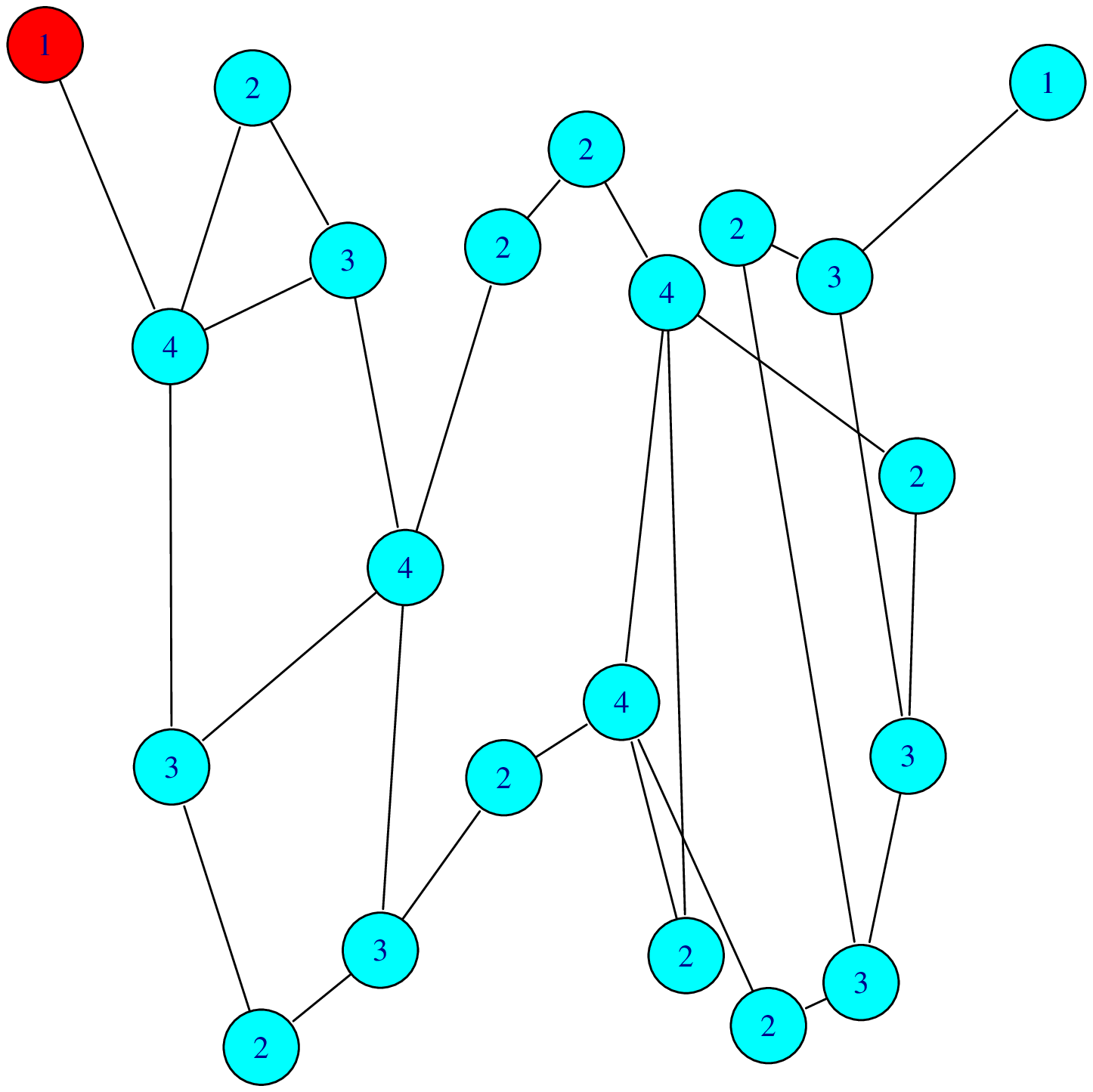}}
\subfigure[D-V-H] {\includegraphics[width = \MySize, angle = -90]
{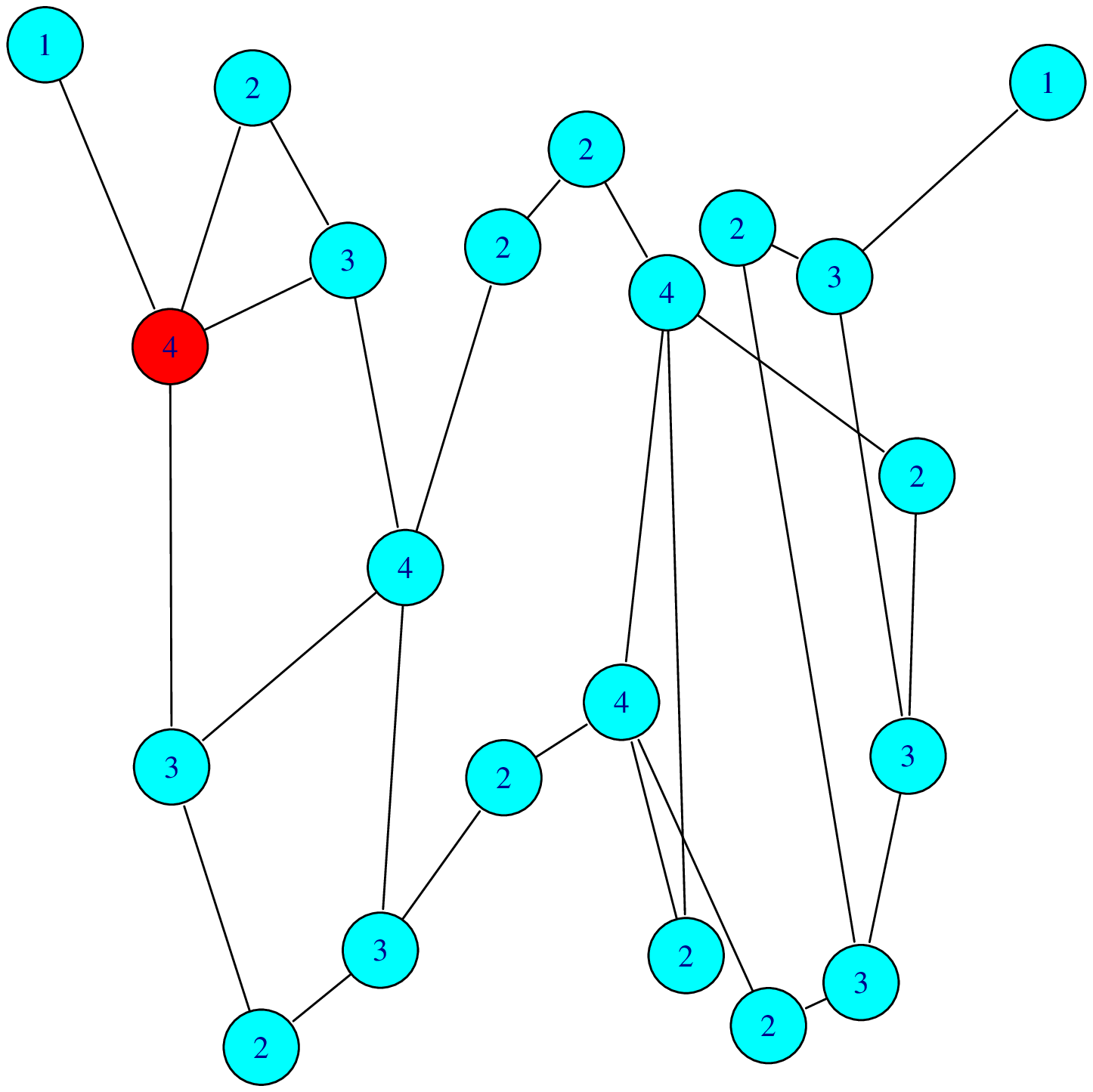}}
\subfigure[C-V-L] {\includegraphics[width = \MySize, angle = -90]
{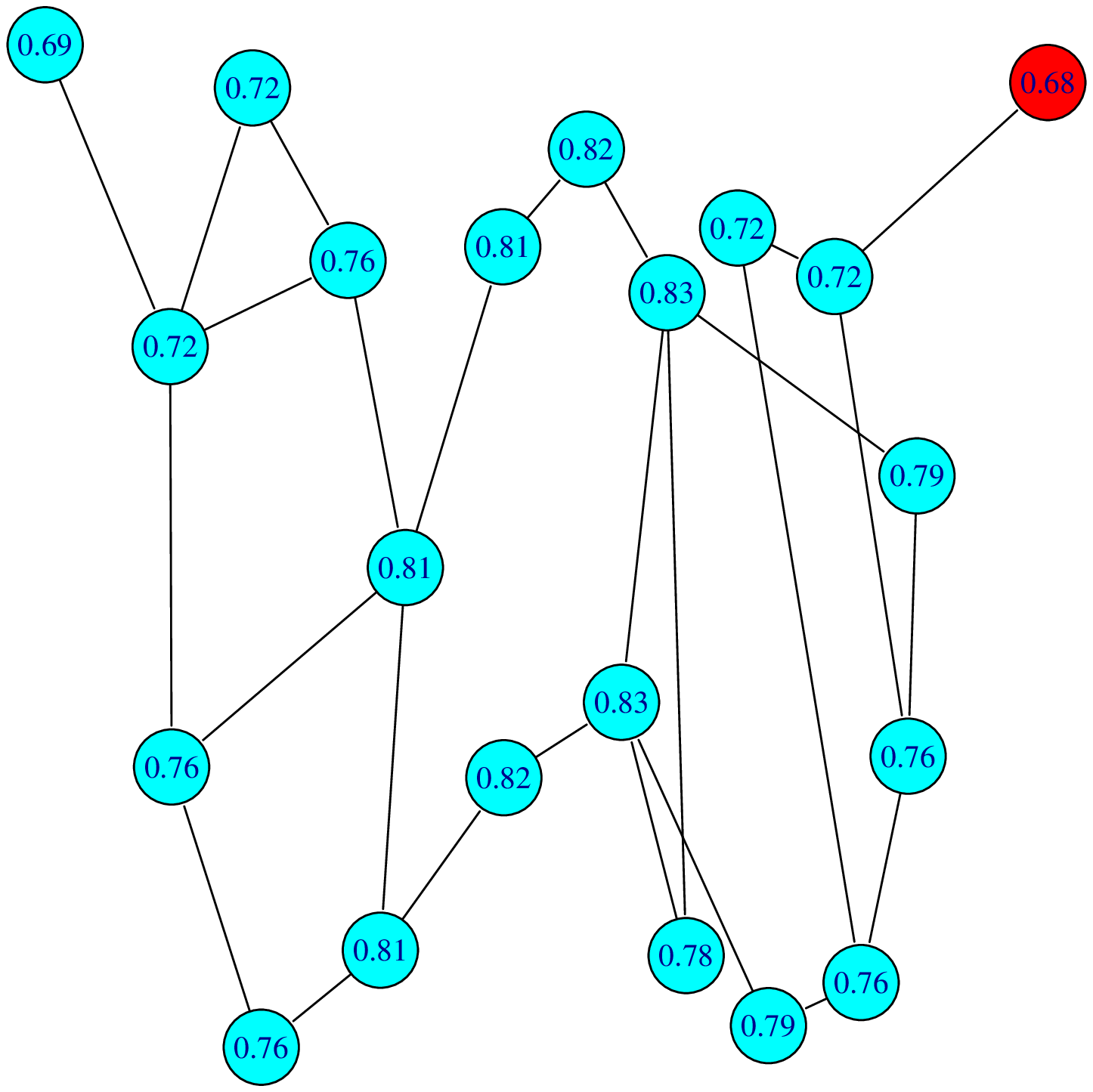}}
\subfigure[C-V-H] {\includegraphics[width = \MySize, angle = -90]
{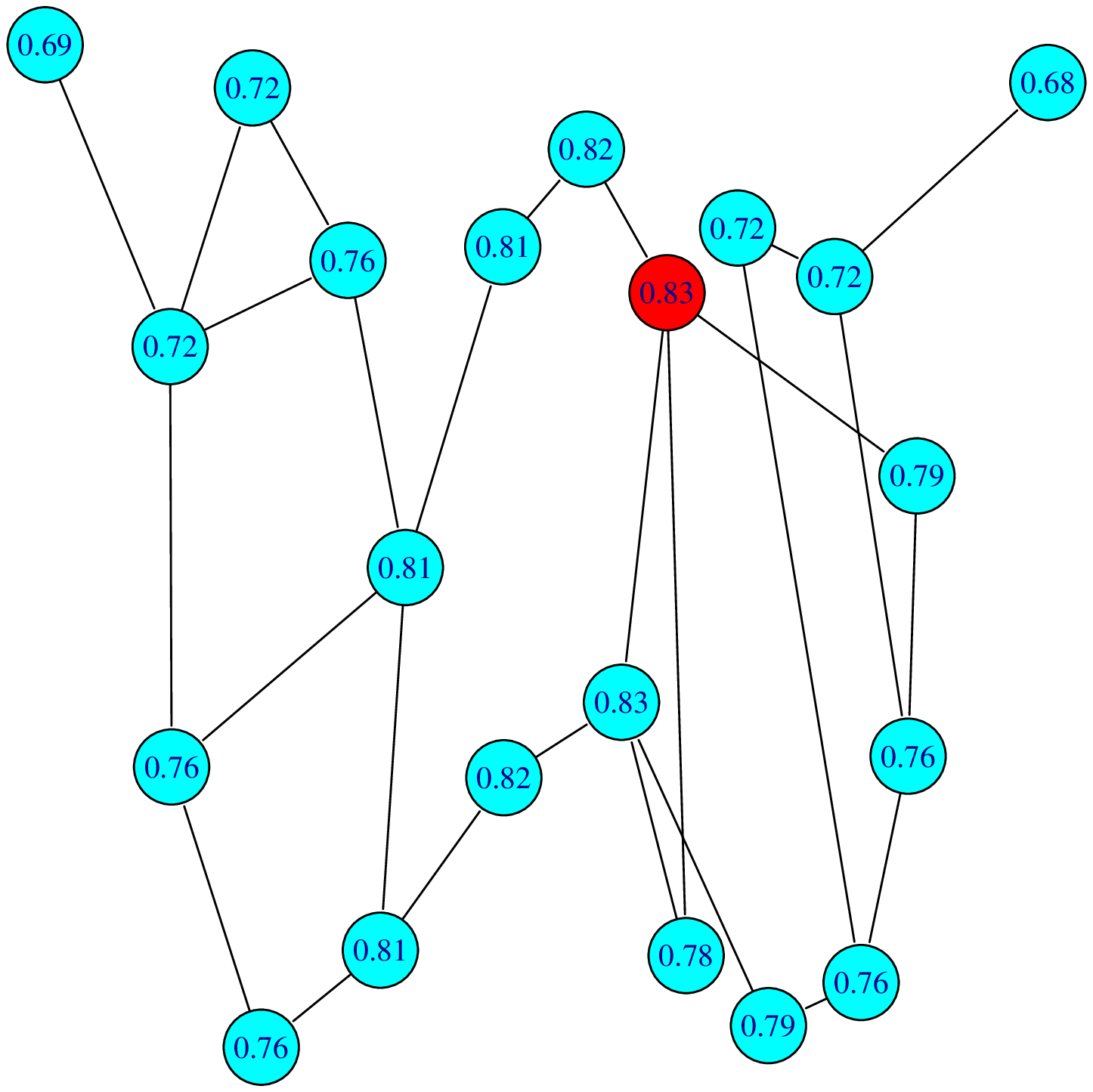}}
\MyCaption{The first graph component that will be removed based on different
attack profiles.}{Each profile selects a different component to be
removed.  In each of these figures, the first component to be removed is
shown in red.  In cases where more than one component has the appropriate qualities to qualify it for removal; selection of which component to remove is based on random selection. }{fig:first}
\end{figure*}

\begin{figure}[t]
\centering
\includegraphics[scale=0.3,angle=-90]{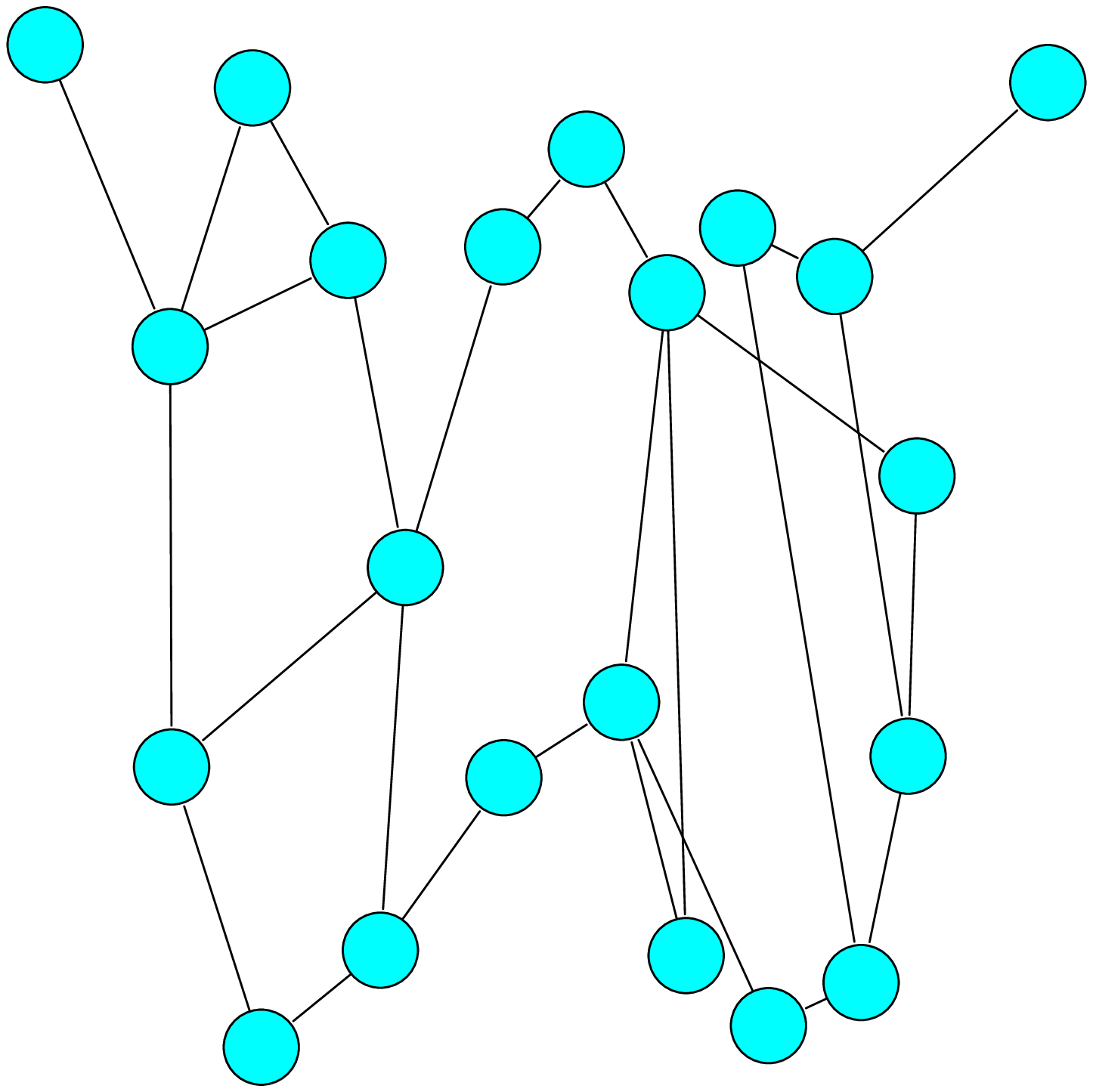}
\MyCaption{The  sample graph presented to each of the attack profile.}{This
graph is used because it has enough interesting vertices and edges so that each
attack profile would select a different graph component to remove.}{fig:sample}
\end{figure}

\subsection{Efficacy of different attack profiles}
Attacker profiles  were used recursively  against the sample graph until the
graph was disconnected.  Table \ref{tbl:efficacy} reports the efficacy of
different attacker profiles used against the sample
graph when  vertices (or edges) are removed.  Computing the centrality value may
result in
more than one vertex (or edge) having the same value.  Having the same value for
two measurements is treated as creating two different graphs that are treated in
a recursive manner.  For each profile, the number of unique graphs is reported
as well as the maximum and minimum recursion depth.  A mean ($\mu$) and standard
deviation ($\sigma$) depths for all profiles are reported.  Smaller values for the
maximum and minimum depths and $\sigma$ points to a profile that is always
aggressive and effective. Good attack profiles (from the attacker's
perspective) are \textbf{D-V-H}, \textbf{B-E-H}, or \textbf{C-V-L} because these
attack the ``core'' elements of the graph, while the other profiles ``nibble at
the edges.''  The best attack profile is  \textbf{B-V-H} because it is most
destructive at the core.  The other attack profiles focus on the periphery and
will result in disconnecting the graph or may result in a connected graph that
has only one vertex.  These peripheral attack profiles take much longer than an
attack on the core.

\begin{table*}[htbp]
\begin{center}
\begin{tabular}{|c|r|r|r|r|r|}
\hline
\textbf{Attack} & \textbf{\# of unique} & \textbf{Max.} & \textbf{Min.} & \textbf{Mean} & \textbf{Std. Dev.}\\
\textbf{profile} & \textbf{graphs} & \textbf{depth} & \textbf{depth} & ($\mu$) \textbf{depth} & ($\sigma$) \textbf{depth}\\
\hline
\textbf{D-V-L} & 428,580 & 20 & 4 & 15.57 & 3.65\\
\hline
\textbf{D-V-H} & 8 & 2 & 1 & 1.87 & 0.35\\
\hline
\textbf{B-E-L} & 7 & 6 & 6 & 6 & 0.00\\
\hline
\textbf{B-E-H} & 2 & 2 & 2 & 2 & 0.00\\
\hline
\textbf{B-V-L} & 53,155 & 20 & 15 & 19.56 & 0.82\\
\hline
\textbf{B-V-H} & 1 & 2 & 2 & 2 & n/a\\
\hline
\textbf{C-V-L} & 2,634 & 20 & 17 & 19.89 & 0.36 \\
\hline
\textbf{C-V-H} & 4 & 2 & 2 & 2 & 0.00 \\
\hline
\end{tabular}
\MyCaption{The  efficacy of different recursive attacker profiles against the
same sample graph.}{}{tbl:efficacy}
\end{center}
\end{table*}

\begin{sloppypar}
\section{Implementing resiliency in graphs of different flavors}\label{append:resiliency}
\end{sloppypar}
The words robustness and resilience are used almost interchangeably when talking about graphs.  But, they are very different attributes that should not be confused.

\emph{Robustness}  \cite{klau2005robustness} is the ability of a graph to keep its basic functionality even under the failure of some of its components.  These components can be any combination of the graph's edges or vertices.  A graph \MyEquationInline{graph} is robust if messages can be sent from \MyEquationInline{allI2J}.

\emph{Resilience} is the ability of a graph to recover readily from damage and, in a sense to become more \emph{robust}.

Implementing resiliency in support of the Game resolved itself into two different categories based on the underlying mechanisms that were used to create the graph.  The ``standard'' graphs (random, power law, and small-world) have a relatively simple underlying mathematical foundation, while the USW graph is created via a series of algorithmic steps.  This division is also reflected in how \emph{resilience} is implemented.

\subsection{Non-unsupervised small-world}
\emph{Resilience} for all ``standard'' graphs is implemented via the use of R igraph library routines.  For all ``standard'' graphs, it is assumed that 10\% of the nodes will be re-activated (via a manner that is outside this discussion) and will attempt to form links to other nodes in the graph.  Further it is assumed that only 90\% of these attempts will be successful.  Connection attempts are not 100\% successful in order to simulate downtime at the host where a node lives and timeouts in the communications channels.
\subsection{Unsupervised Small World}
Unsupervised Small World (USW) graphs are created using a number of control parameters.  These parameters autonomously create a graph that has small-world characteristics using only locally gained knowledge.  The parameters that were used to create the graph are the same ones that are used to implement resilience.

During the simulation, when the USW graph is written to the database, all the control parameters ($\alpha, \beta, \gamma$ and others) are also written.  When the USW graph is to be reconstituted, the original control parameters are used to reconstitute the graph.

\section{The Game}
A \emph{Game} is a competitive activity involving skill, chance, or endurance
between two or more  players who play according to a set of rules to achieve
some goal.  There are two players in this game: Alice (the person responsible
for repairing the graph) and Mallory (the attacker).

A graph may be subject to different types of damage.  Damage resulting from
a random event or occurrence  can be classified as is an error
 \cite{Albert2000}.  While, damage from something other  than a random act, is
classified as an attack.  For example, loss of a single router in a computer
network could be viewed as a random event.  Loss of all routers at the same time
would be called an attack.  Structurally different types of graphs (random,
power law, small-world)  \cite{lee2006rnt,
motter2002cascade, criado2005effective, tanizawa2005optimization, zio-modeling,
klau2005robustness,
callaway2000network, netotea2006evolution, bollobas2004robustness}
are robust in different ways when the same number of graph
components are lost.

Albert, Jeong and Barab{\'a}si  \cite{Albert2000} focus on power law, networks
such as the World-Wide Web, the Internet, social networks and cells.  They
conclude that these networks have are tolerant to many random failures, but are
very susceptible to the failure of a few critical elements because of their
underlying structure.  This type of sensitivity is common to power law
networks.  USW is not a power law graph and is not sensitive to targeted
attacks.

Moreno, Pastor-Satorras et al,  \cite{moreno2003critical} focus on the effects
of a cascading failure in a power law network.  Using their analysis, they
identified a critical load in the traffic through a failed network component
above which the resulting traffic congestion will destroy network
communications.  This critical threshold is based on the idea that each
component has a communications tolerance that when exceeded causes the
component to fail.  By keeping these limitations in mind, USW graphs have
been designed without tolerances and are able to send as much traffic as the
underlying Web architecture can support.  Zio and Sansavini  \cite{zio-modeling}
expand on the ideas of Moreno and  Pastor-Satorras by exploring small-world
graphs as well.  Zio and Sansavini go on to quantify the amount of excess
capacity (of a normalized loading) that each node in both types of graphs must
have to prevent cascading failures.

Guillaume, Latapy and Magnien  \cite{guillaume2005comparison} extend Albert,
Jeong and Barab{\'a}si investigation to include random graphs.  They show that
the removal of a similar number of edges for the two graph types will result in
a disconnected graph, and then propose an efficient attack strategy
based on removal of edges.  The attack ideas were incorporated into the attack
profiles used to test the robustness and resilience of the USW graphs.

Motter and Lai  \cite{motter2002cascade} focus on the effects of cascading
failures due to overloading of the Internet and power grids.  In these types of
graphs the traffic when a component
fails, the traffic (be it either packets or electrical power) being serviced by that component is transfered
to other components of the same type to which the failed component was
connected.  Their analysis shows that an attack, or a failure of an
exceptionally heavily loaded component may have a cascading failure affect on
other components.  Traffic between USW components are not bound or limited
except by the underlying Web Architecture and are those immune from these types
of cascading failures.

Farkas, Antal, et al.  \cite{farkas2005dra} proposed imbuing an ethernet
network with a distributed resilient architecture based on the use of
multiple static routes stored in routers spread across the network.  In the
event that a router were to fail, or become unavailable, the connected routers
would immediately begin using the secondary spanning tree routes.  These
secondary routes would be maintained in addition to the primary routes.  The
USW model does not maintain static or dynamic routes.

Criado, Flores, et al.  \cite{criado2005effective} propose two measures to
assess the robustness of a graph to random and intentional attacks.  These
measures take into account the graph's topology and is computable in polynomial
time.  Their measures can be viewed as another type of centrality where the
node (or edge) whose presence means that the graph is less vulnerable and whose
absence wold make the graph more vulnerable.  The evaluation of the
vulnerability of USW  graphs will take their ideas into consideration.

Netotea and Pongor \cite{netotea2006evolution} take as input a graph and through
evolution increase it's efficiency and robustness by rewiring
the graph.  They take existing an existing edge and move one end to a different
node and then measuring the efficiency of the graph at each stage.  Their definition
of efficiency is:
\MyEquationLabeled{globalEfficiency}{equ:globalEfficiency} whereby the
average distance between all nodes decreases.  They define robustness at time
$t$ as  \MyEquationInline{globalRobustness}.  Within USW, once edges are
created, they are not removed or altered.  The efficiency of the USW graph will
increase by the addition of more nodes and edges.

\subsection{The goal}
Our goal for the game is to determine the \emph{robustness}  of selected graph
types in the face of different types of attacks (\emph{attacker profiles}) and
how \emph{resilient} each graph type is when given an opportunity to
recover from  some of the damage suffered in the attack.
The game seeks to answer the question: how many edges or  nodes can the attacker
remove before the graph was disconnected?  Different attack profiles will be
exercised and the worst type of attack (i.e., the one with the highest
likelihood of success from the attacker's perspective) will be identified.
Each of the graphs described in Section \ref{sec:typeOfGraphs} will compete
against a set of attack profiles.  Each attack will use a different and
unchanging  profile against the graph.  The game is over when either: the
graph is disconnected, or the simulation runs to an end.
\subsection{The players}
During the game; Mallory will have global knowledge of the graph (how this
knowledge was obtained is not part of the game) and can choose to remove
any graph component (either edge or
vertex) that he feels is to his benefit. Alice has only local knowledge and
does not know how or where the next  attack will occur. After the
Mallory's turn, Alice will have a turn to reconstitute the graph  in preparation
for the next attack.  Mallory and Alice will alternate turns until one wins.
The goal of the game from the Mallory's perspective is to cause the graph to
become disconnected and therefore Alice would not be able to use the graph
to send a message to
Bob.  The goal from the Alice's perspective is to remain connected as long
as possible.
\subsection{The rules}
There are few rules in this game.  They are:
\begin{enumerate}
 \item The graph is created without any interference from Mallory.
 \item Once  Mallory chooses an attack profile, he must use that same profile for the
duration of the game.
 \item The  number (or percentage) of graph components that Mallory can damage
per turn is fixed at the start of the game.
 \item The  number (or percentage) of graph components that Alice can
reconstitute per turn remains fixed for the duration of the game.
 \item The  game is over if the graph is disconnected at the end of Mallory's
turn, or the number of game turns reaches the maximum number allowed.
\end{enumerate}
If at the  end of the game the graph is disconnected then Mallory has won,
otherwise Alice has won.
\subsection{A sample game}
Data are collected and analyzed during the course of the game between Mallory
and Alice.  A specific instance of a 10 turn game is shown in Figures
\ref{fig:p-2-shots} and \ref{fig:p-2-rounds}.  A power law degree distribution
graph was created and a \textbf{D-V-H} attack profile was used for
10 turns against the graph.  As can be seen in Figure \ref
{fig:p-2-shots} the induced subgraph was severely damaged during the first 3
turns and then less damaged later.    Two of nine turns
are highlighted.  During turn 2, the damaged induced by vertex
deletion increases during the turn and the overall damage to the
graph is shown at the end of the turn.  During turn 4, there is
minimal damage to the subgraph and the overall damage to the
total graph is low.  The worst damage was done to graph be the
end of the fourth turn, after which the graph had reconstituted itself enough to
withstand future attacks.

\renewcommand{\MySize}[0]{3.90in}

\begin{figure*}
 \centering
\includegraphics[width=\MySize,
angle=-90]{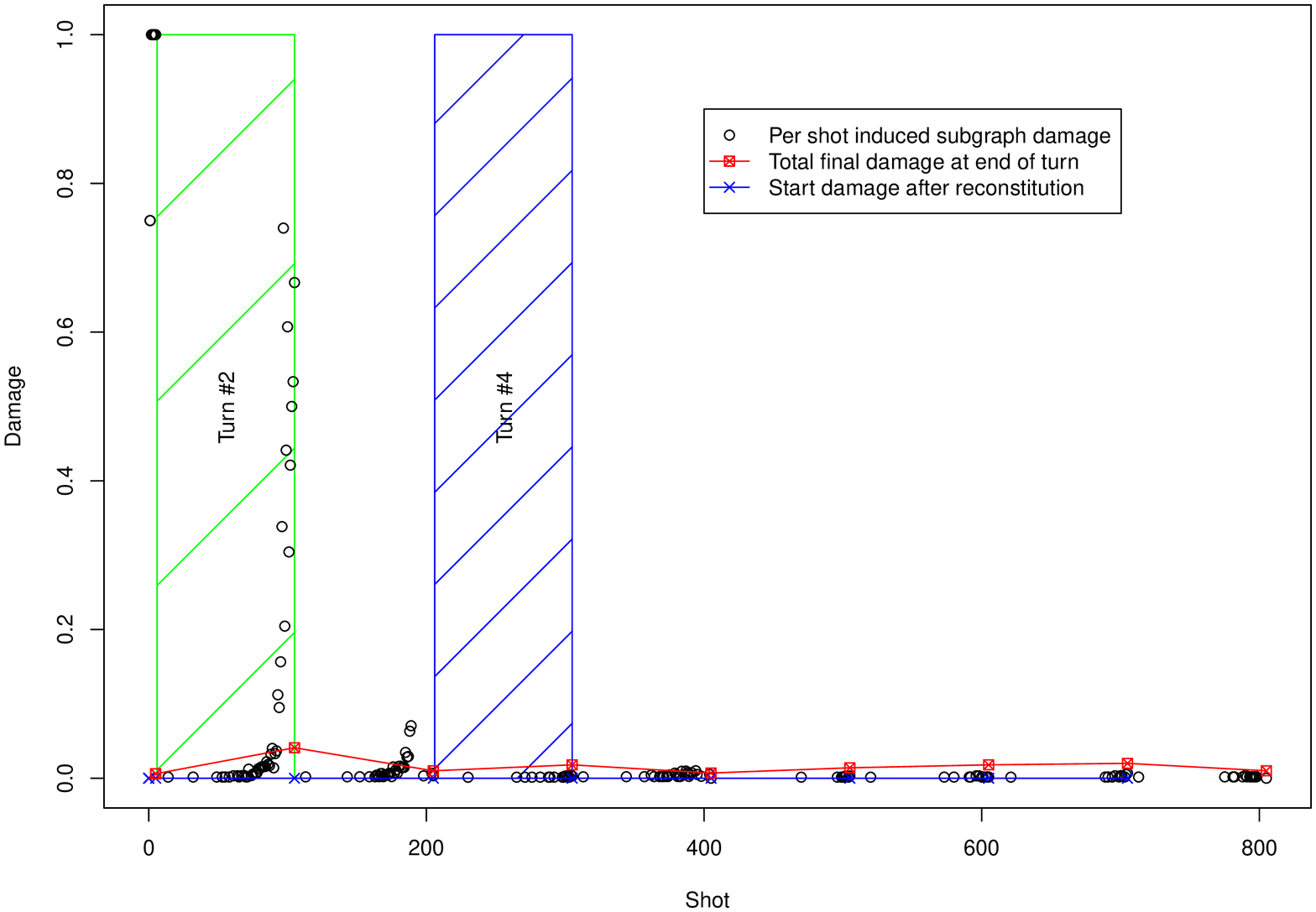}
\MyCaption{Deletion (shot) by deletion plot of damage to a power law degree graph.}{The red line presents the
damage to the graph at the end of a turn.  Each marker on the red line
indicates the end of a turn.  The  blue line is the damage at the start of a
turn. The black markers represent the damage done to an induced subgraph where
the path length (PL) from the root node is 2.}{fig:p-2-shots}
\end{figure*}

\begin{figure*}
 \centering
\includegraphics[width=\MySize,
angle=-90]{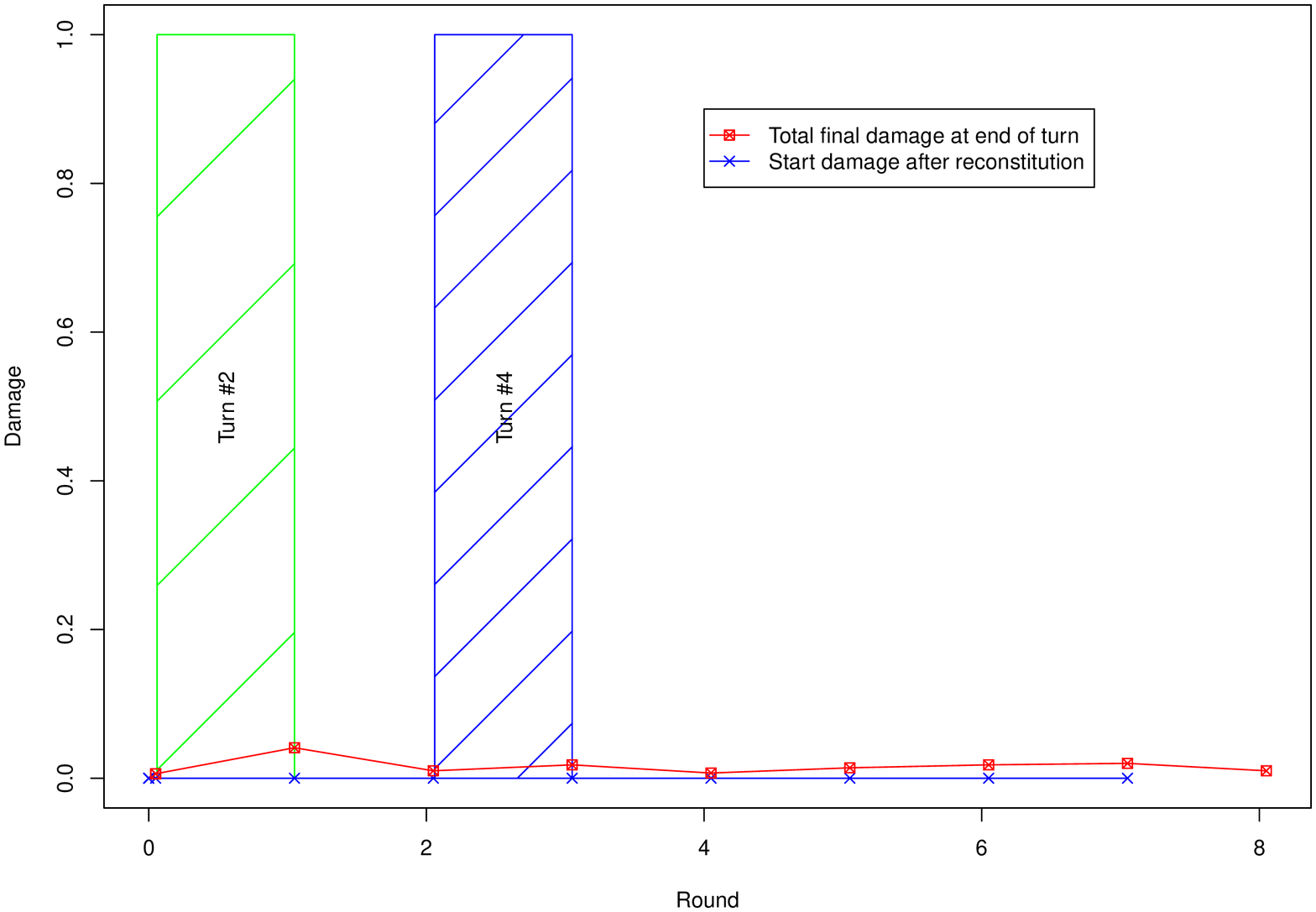}
\MyCaption{Turn by turn plot of damage to a power law degree distribution graph.}{The red line presents the
damage to the graph at the end of a turn.  The  blue line is the damage at the
start of a turn.}{fig:p-2-rounds}
\end{figure*}

\section{Graphs Taken to Disconnection}

\subsection{What data was collected}
At the end of each turn, the following measurements or characteristics are
collected and presented:
average path, clustering coefficient, density and diameter.

The y-axis is normalized from 0 to 1 for all figures. The x-axis
is linear and represents the ``shot'' taken by the Mallory at the graph.
Mallory had 100 shots (10\% of a 1000 vertex graph) per turn.
\subsection{How graphs were created}
 The R igraph package \cite{csardi2006igraph} was used to create the
``standard'' graphs from Section \ref{sec:typeOfGraphs}.  The USW graph creator
program was set up to run with all possible combinations of first attachment,
node visitation and queue processing policies and 4 values each for $\beta$ and
$\gamma$.  Every graph was attacked based on degreeness,
betweenness and closeness for either edge or node as appropriate and for each
possible extremal value
{%
\newcommand{\mc}[3]{\multicolumn{#1}{#2}{#3}}
\begin{table*}[t]
\begin{center}
\scalebox{0.8} {
\begin{tabular}[t]{|l|l|l|l|l|l|l|l|l|}
\hline
& \mc{2}{|c|}{\textbf{Degreeness}} & \mc{4}{|c|}{\textbf{Betweenness}} & \mc{2}{|c|}{\textbf{Closeness}} \\
\cline{4-7}
 & \mc{2}{|l|}{} & \mc{2}{|c|}{Edge} & \mc{2}{|c|}{Node} & \mc{2}{|l|}{} \\
 \cline{2-9}
 \textbf{Graph type}& Lowest &  Highest & Lowest & Highest & Lowest & Highest & Lowest & Highest \\
\hline
Random (Erdos --- Renyi)& 99 & 31 & 344 & 6 & 99 & 14 & 99 & 23 \\
\hline
Power law(Barabasi) & 99 & 1 & 1 & 1 & 99 & 1 & 99 & 1 \\
\hline
Small-world (Watts --- Strogatz)& 99 & 65 & 355 & 85 & 99 & 10 & 99 & 21 \\
\hline
USW & 99 & 65 --- 99 & 1100 --- 31000 & 13 --- 60 & 99 & 35 --- 99 & 99 & 70 --- 90 \\
\hline
\end{tabular}
}
\MyCaption{A comparison of how many graph elements (either edges or vertices) must be removed before the graph becomes disconnected.}{}{tbl:results}
\end{center}
\end{table*}
}%
\begin{table*}
\begin{center}
\begin{tabular}{|l|p{5in}|}
\hline
\textbf{Attack profile} & \textbf{Efficacy this is the one}\\
\hline
[BD][EV]L & ``Nibbles'' away at the least important component, gnawing at the
graph until a critical component (or the last node) is reached.\\
\hline
[BD][EV]H & Attacks the most critical/needed part of the  graph and is able to
disconnect the graph after a relatively few number of removals.\\
\hline
CVL & Removes the node that is ``closest'' to all nodes and  forcing another to
become the closest.\\
\hline
CVH & ``Nibbles'' away at the periphery (in a closeness sense) of the graph and
continues to do so until there is only one component left or a disconnection.\\
\hline
\end{tabular}
\MyCaption{A summary of efficacy of different attack profiles.}{}{tbl:efficacyComments}
\end{center}
\end{table*}

\subsubsection{Non-unsupervised small-world}
\begin{sloppypar}
The R package igraph functions erdos.renyi.game$()$, barabasi.game$()$ and
watts.strogatz.game$()$ were used to create random, power law and
small-world graphs respectively of 1000 nodes each. Each graph was checked to
ensure that it was simple and connected.
\end{sloppypar}
\subsubsection{Unsupervised Small World graphs}
A complete description of the  USW construction process can be found in
\cite{jcdl09-usw,1378990}.  The dominant control parameters for the
destruction game are:
\begin{description}
 \item  [b ($\beta$)]  the threshold that a locally generated random number must
exceed before an edge can be created between two nodes
 \item  [p ($\gamma$)] the percentage of nodes that failed or have never been
considered for a connecting edge that will be used once the $\beta$ threshold
test has been satisfied
 \item  [t] the percentage of the graph selected for
re-constitution to support resiliency corrective actions will attempt to connect
to using both the initial $\beta$ and $\gamma$ values
 \item [L] the number of nodes that will be randomly selected for re-constitution actions
\end{description}

Table \ref{tbl:results} itemizes the results of the various attack profiles against the ``standard'' and USW graphs.
\section{Discussion}
Mallory is in a much stronger position than Alice.  Mallory's knowledge of the
graph is only limited by the amount of time and energy that he wishes to expend
exploring the graph.  If the depth of the graph that Mallory wishes to explore
is called path length (PL), then the range on the size of the induced subgraph he will
discover could grow exponentially based on PL \MyEquationReference{equ:inducedSize}.

\MyEquationLabeled{graphSize}{equ:inducedSize}

Where \MyEquationInline{averageDegreeRandom} is the average
number of edges adjacent to any node in the graph.

Even modest increases in $PL$ (for example
from 2 to 5) can have profound effect on the size of the induced subgraph.  By way of illustration, it
has been estimated that the entire Internet is no more that 19 clicks (edges or
PL)
in size \cite{barabasi2001physics}.

Alice is a severe disadvantage.  She is limited to purely local knowledge (i.e., PL =
1), and can only randomly select which nodes to use to reconstitute the graph.
The percentage of vertices that she can activate has to be high enough to have
a reasonable expectation of overlapping those graph components that Mallory
has attacked.  If Alice can cover those components that Mallory has
affected then the graph will continue to survive, otherwise it is inevitable
that the graph will become disconnected.

Data from Table \ref{tbl:results}  was used to bound the region  where the USW
graphs are most vulnerable (most vulnerable means that the attacker can remove
the fewest elements and disconnect the graph).
Table \ref{tbl:efficacyComments} summarizes the comments  in the previous
subsections about the efficacy of the different attack profiles.
After evaluating the efficacy of the various attacker  profiles; \textbf{B-V-H}
was selected as best (from the attacker's perspective).

\section{Conclusions}
We have demonstrated that a graph of WOs created
based on the USW algorithm is both robust and resilient in a simulation
environment.  The next major step will be to take it from the theoretical
environment and to a real-world implementation.  One possible implementation
is to further the ideas in  \cite{mccown:everyone_is_a_curator} by writing
and fielding a Firefox plugin using the Open Archives Initiative (OAI) Object
Reuse and Exchange (ORE) model from  \cite{ore:techreport:2008} to
identify data that could be submitted to a digital library for preservation.
These WOs would be monitored for some length of time in order to see how well
reality matches the simulations.

We investigated  the efficacy of various attack profiles on a graph of web
objects.  The most effective (i.e., the most destructive) is when the attacker
uses a \textbf{B}etweenness \textbf{V}ertex \textbf{H}igh centrality
measurements to select which vertex to remove.
 The size of the subgraph that the attacker can focus on is dependent on the graph's average degree connectivity
\MyEquationInline{degreeConnectivity}  and is exponential on the Path Length
from a root node.  If the number of graph components an attacker can remove is
greater than the number that can be reconstituted then the graph will eventually
be destroyed.  Based on simulations, the Unsupervised Small World (USW) graphs
are more robust and resilient
 then those constructed using
classical random, power law, or Watt---Strogatz techniques.  A graph of USW connected WOs filled with data
could outlive the individuals and institutions that created
the data even in an environment where WOs are lost due to random failures or directed attacks.
\section{Acknowledgment}
 This work supported in part by the NSF, Project 370161.  We also wish to thank the three anonymous
referees for their constructive comments on an earlier version of this paper.
\bibliographystyle{abbrv}
\bibliography{local}  
\end{document}